\documentclass[11pt]{article}

\usepackage{graphicx}
\usepackage{amsmath}
\usepackage{amssymb}
\usepackage{pslatex}
\usepackage{color}
\allowdisplaybreaks[3]

\makeatletter

\@addtoreset{equation}{section}
\makeatother

\begin{document}
\begin{titlepage}
\mbox{}
\vspace{25mm}
\begin{center}
{\Large\bf%
Classical Solutions of a Torsion Gravity from a Large $N$ Matrix Model 
}\vspace*{10mm}

Hiroshi Isono\footnote{e-mail: 
{\tt isono@ntu.phys.edu.tw }}
and
Dan Tomino\footnote{e-mail: 
{\tt tomino@phys.cts.nthu.edu.tw}}
\vspace*{5mm}

${}^1$
{\it Department of Physics and Center for Theoretical Sciences, National Taiwan
University, Taipei 106, Taiwan 
\\[3mm]
${}^2$
Physics Division, National Center for Theoretical Sciences, Hsinchu 300, Taiwan
}\\[1mm]

\end{center}

\vspace*{10mm}
\begin{abstract}\noindent%
Large $N$ matrices can describe covariant derivatives in curved space.
Applying this interpretation to the IKKT matrix model, the field equation of gravity is derived from the matrix equation of motion. We study classical solutions of this field equation with torsion degrees of freedom in empty spacetime. 
Time dependent solutions with homogeneity and isotropy, and time independent solutions with spherical symmetry are investigated under particular settings of torsions.     
\end{abstract}

\end{titlepage}

\newpage

\section{Introduction}
In matrix model proposals for type I{}IB superstring theory and M-theory \cite{Ishibashi:1996xs,Banks:1996vh}, describing gravity by matrices is one of the important subjects. Gravity is not prepared at formulations of these matrix models, but it must be encoded in their matrix degrees of freedom. Several years ago, a new interpretation of large $N$ matrix was proposed by Hanada, Kawai and Kimura \cite{Hanada:2005vr,Kawai:2007zz}; large $N$ matrices can describe covariant derivatives in curved space. They applied this new interpretation (we call this interpretation ``HKK interpretation" or simply ``HKK" for later convenience) to the IKKT matrix model \cite{Ishibashi:1996xs}, and derived the  Einstein equation in empty space from the classical matrix equation of motion. Several studies have been done to explore this attractive proposal \cite{Hanada:2006gg,Saitou:2006ca,Hanada:2006ei,Furuta:2006kk,Matsuo:2008yd}\footnote{Other similar attempts based on non-commutativity can be found in \cite{Steinacker:2007dq}}.

In the HKK interpretation, large $N$ matrices contain not only vielbein and spin connection
but also infinitely many other degrees of freedom with various higher spins.
If one consider gravity derived by HKK without these higher spin fields, 
this is  regarded as an effective theory in nearly empty space.   
Even under such simplification, however, this effective theory has more degrees of freedom than Einstein gravity.   
In HKK, the vielbein and the spin connection encoded in large $N$ matrices are independent variables,
then the matrix description of gravity gives rise to torsion degrees of freedom unless we impose some torsion free conditions by hand.
Therefore IKKT matrix model provides a torsion gravity.
This arouses our curiosity on the role of torsion in the gravity from the matrix model. The classical field equation would be a good way to capture its semi-classical dynamics. 
Gravity equations with torsion from several bosonic IKKT-type matrix models have been written down in \cite{Furuta:2006kk}.
Our purpose in this paper is to investigate classical solutions of the gravity equation with torsion, from the bosonic IKKT model, under simple settings of torsion. \\

There are two remarks:
\begin{itemize}
\item A generalization of Einstein gravity to including torsion  has been known as the Poincar\'{e} gauge theory of gravity (PGT), and has been studied extensively (see \cite{Hehl:1976kj, Hammond:2002rm} for reviews). 
It is described by a gravity Lagrangian which includes bi-linear terms of torsion. Vierbein and spin connection in this Lagrangian are independent variables.  
On the other hand, in this paper we will study a gravity equation with torsion which is derived from the HKK interpretation of a matrix model equation of motion, which is different from what is derived from the PGT.    
Although the matrix equation of motion is derived from a matrix model action, 
the gravity equation of HKK seems difficult to derive by a variation of any gravitational Lagrangian.

\item In \cite{Furuta:2006kk}, parts of torsion degrees of freedom were identified with scalar and 2nd rank antisymmetric tensor fields,
which are analogous to the massless fields of string theory. 
In this paper we do not always follow such an interpretation but rather keep broader possibility for solutions.  
\end{itemize}

The organization of this paper is as follows:
In section 2, we briefly describe the gravity equation with torsion from the HKK interpretation and fix our notation. The original proposal was done for a matrix model with an Euclidean signature. After writing down the field equation with Euclidean signature, we rotate it to one with a Lorentzian signature.    
In section 3, we consider time dependent solutions under homogeneous-isotropic ansatz. 
In section 4, static and spherical symmetric solutions are studied.
In addition to the analytic method, we examine numerical computation to find  solutions.
Finally we summarize and discuss our results in section 5.

\section{Gravity Equation from Large $N$ Matrix Model}
Large $N$ matrices can describe covariant derivatives in curved spaces \cite{Hanada:2005vr,Kawai:2007zz}. In this interpretation, the large $N$ matrices are regarded as a map between sections of a fiber bundle over a curved manifold whose fiber is the vector space of structure group ${\rm Spin}(d)$. Formally differential operators may be expanded by a power series of covariant derivatives and Lorentz generators. Associated with this expansion, various fields: ${\rm U}(1)$ gauge field, vielbein, spin connection and other infinitely many higher spin fields appear. Here we concentrate ourselves on the degrees of freedom associated with the vielbein and spin connection. In this situation the HKK interpretation is simply stated as
\begin{eqnarray}
A_{(a)}=R_{(a)}{}^a(g^{-1})\nabla_a\,,
\label{HKK}
\end{eqnarray}       
where index $(a)=1,2, ... ,d$ labels $N\times N$ Hermitian matrices, on the other hand, another index $a$ is the local Lorentz indices in $d$-dimensional curved manifold.
${R_{(a)}}^a(g^{-1})$ is in the vector representation of the local Lorentz group ${\rm Spin}(d)$ whose elements are denoted by $g$. We describe curved space covariant derivative $\nabla_b$ by the vielbein $e^{\mu}{}_a$ and the spin connection 
$\omega_{a}{}^{bc}$ as 
\begin{eqnarray}
\nabla_a=e^{\mu}{}_a\partial_{\mu}+\omega_{a}{}^{bc}{\cal O}_{bc},
\end{eqnarray}  
where $\mu$ is the curved space index and ${\cal O}_{bc}$ is the Lorentz generator.

Commutator of these covariant derivatives gives
\begin{eqnarray}
[\nabla_a, \nabla_b]=-T^c{}_{,ab}\nabla_c+R_{ab}{}^{cd}{\cal O}_{cd}.
\end{eqnarray}
Here $T^c{}_{,ab}$ and $R_{ab}{}^{cd}$ are defended as
\begin{eqnarray}
-T^{c}{}_{,ab}&=&e_{\mu}{}^{c}(\partial_{a}e^{\mu}{}_{b}-\partial_{b}e^{\mu}{}_{a})+\omega_{ab}{}^c-\omega_{ba}{}^c, \nonumber \\ 
R_{ab}{}^{cd}&=& \partial_{a}\omega_{b}{}^{cd} - \partial_{b}\omega_{c}{}^{cd}+\omega_{a}{}^c{}_{e}\omega_{b}{}^{d}{}_{e}-\omega_{b}{}^c{}_{e}\omega_{a}{}^{d}{}_{e}
\label{defofTR}
\end{eqnarray}
where $\partial_a=e^{\mu}{}_a\partial_{\mu}$. We do not impose torsion free condition by hand. Thus the torsion $T^c{}_{,ab}$ is non-vanishing and $e^{\mu}{}_a$ and $\omega_{a}{}^{bc}$ are independent of each other.

Next, to obtain a gravity equation, we apply this interpretation to the IKKT-type matrix model ($d=10$ corresponds to the IKKT model). The bosonic part of the matrix model action is 
\begin{eqnarray}
-\frac{1}{4}{\rm Tr}[A_{(a)},A_{(b)}][A_{(c)},A_{(d)}]\delta^{(a)(c)}\delta^{(b)(d)}.
\end{eqnarray} 
Matrix equation of motion from this action becomes
\begin{eqnarray}
[A^{(a)},[A_{(a)},A_{(b)}]]=0.
\label{mateqm}
\end{eqnarray}
We now apply the HKK interpretation (\ref{HKK}) to (\ref{mateqm}). With a help of a property (68) in \cite{Hanada:2005vr}, this equation becomes
\begin{eqnarray}
[\nabla^a, [\nabla_a,\nabla_b]]=0.
\end{eqnarray}
From this equation we obtain following equations
\begin{eqnarray}
R_{ab}+\nabla^cT_{b,ca}+T^{p,q}{}_aT_{b,pq}=0, \nonumber \\
\nabla^aR_{abcd}+T^{p,q}{}_bR_{pqcd}=0,
\label{fieldequation}
\end{eqnarray}
where $R_{ab}=R_{aeb}{}^{e}$. 

From a matrix Jacobi identity
\begin{eqnarray*}
[\nabla_{(a)}, [\nabla_{(b)},\nabla_{(c)}]]
+[\nabla_{(b)},[\nabla_{(c)},\nabla_{(a)}]]
+[\nabla_{(c)},[\nabla_{(a)},\nabla_{(b)}]]=0,
\end{eqnarray*}
we have another set of equations:
\begin{eqnarray}
-\nabla_{(a}{T^d}_{,bc)}-{R_{(bca)}}^d+{T^e}_{(,bc}{T^d}_{,a)e}=0, \nonumber \\
\nabla_{(a}{{R_{bc)}}^d}_e-{T^f}_{,(ab}{{R_{c)f}}^d}_e=0. 
\label{jacobi}
\end{eqnarray}
These equations are identical with Bianchi identities of ${T^c}_{,ab}$ and ${R_{ab}}^{cd}$. Thus (\ref{jacobi}) is automatically satisfied according to the definitions (\ref{defofTR}). See the Appendix as a proof.  \\

Equations (\ref{fieldequation}) are derived from IKKT-type matrix model with Euclidean signature. Having a gravity equation once, we rotate it to have Lorentzian signature in order to discuss classical solutions which describe manifolds with Lorentzian signature. Then, (\ref{fieldequation}) with Minkowski signature are basic equations which we will use from the next section.

\section{Time Dependent Solutions with Homogeneity and Isotropy}
In this section we consider time dependent solutions with assumption of homogeneity and isotropy. General ansatz for the spin connection under this assumption was already proposed in PGT \cite{Gonner:1984rw}. Here we do not use this most general ansatz. Rather we shall consider simple settings as possible, and examine the role of the torsion in our classical gravity.

\subsection{Time dependent solutions with single scale factor}  
We shall consider $(n+1)$-dimensional homogeneous and isotropic spacetime. For the metric 
and the spin connection we adapt following ansatz:
\begin{eqnarray}
&&ds^2=-dt^2+a(t)^2\sum_{i=1}^{n}dx_i^2,  \nonumber \\
&&{\omega_{1}}^{1}{}_{0}= {\omega_{2}}^{2}{}_{0}= \cdots ={\omega_{n}}^{n}{}_{0}\equiv X(t), \quad
\mbox{(others)}=0. 
\end{eqnarray}
Under this ansatz non-vanishing components of torsion are
\begin{eqnarray}
T^{i}{}_{,i0}=X-H, \quad \left(H\equiv\frac{\dot{a}}{a}\right). 
\end{eqnarray} 
Substituting these forms into (\ref{fieldequation}), we obtain following equations:
\begin{eqnarray}
\dot{X}+X^2=0, \label{feq31} \\
2\dot{X}-\dot{H}+(2n-1)X^2-nHX+H^2=0, \label{feq32} \\
(\partial_t+nX-H)(\dot{X}+HX)-(n-1)X^3=0, \label{feq33}
\end{eqnarray}
where (\ref{feq31}) and (\ref{feq32}) come from the first equation of (\ref{fieldequation}), and (\ref{feq33}) comes from the second equation of (\ref{fieldequation}). We have three equations here, but it turns out that (\ref{feq33}) can be written using (\ref{feq31}) and (\ref{feq32}). 
Therefore we consider (\ref{feq31}) and (\ref{feq32}) as independent equations to determine $X$ and $H$.\\

First, we find  a solution of (\ref{feq31})-(\ref{feq33}) with following form:
\begin{eqnarray}
X=0, \quad H=\frac{1}{\alpha-t},
\label{sol31}
\end{eqnarray} 
with an integral constant $\alpha$. 
The scale factor becomes $a(t)\propto (\alpha-t)^{-1}$, so it blows up as time with positive acceleration i.e. $\ddot{a}>0$.
In this solution all quantities calculated by Riemann curvatures vanish because $X$ is zero.   
Expansion of the space by $a(t)$ is purely supported by the torsion.

Next we consider $X\neq 0$ solution. 
We can solve (\ref{feq31}) by $X=(\beta+t)^{-1}$ where $\beta$ is an integral constant.
Substitute it into (\ref{feq32}), we have
\begin{eqnarray}
\dot{H}=H^2-\frac{n}{\beta+t}H+\frac{2n-3}{(\beta+t)^2}.
\label{feq34}
\end{eqnarray}
A solution of (\ref{feq34}) can be written by
\begin{eqnarray}
H=\frac{u}{\beta+t},
\end{eqnarray}
and $u$  is determined by second order equation
\begin{eqnarray}
u^2-(n-1)u+(2n-3)=0.
\end{eqnarray}   
There are real solutions of $u$ if $n=1$ or $n\ge 9$. 
The scale factor becomes $a(t)\propto (t+\beta)^u$ and $\ddot{a}>0$ is possible if $n\ge 9$.

In the case of $1<n<9$, there is no simple analytic solution. However, (\ref{feq34}) tells us that its solution approaches to (\ref{sol31}) as $t$ increases. 
Thus finally it blows up with positive acceleration. \\

Every three cases which we have considered implys that introducing torsion gives expansion with positive acceleration to the universe.

\subsection{Time dependent solutions with two scale factors}
We shall consider $(n+m+1)$-dimensional homogeneous and isotropic spacetime with two different scale factors. For metric 
and spin connections we adapt the following ansatz:
\begin{eqnarray}
&&ds^2=-dt^2+a(t)^2\sum_{i=1}^{n}dx_i^2+b(t)^2\sum_{I=n+1}^{n+m}dx_I^2,  
\nonumber \\ 
&&\omega_{1}{}^{1}{}_{0}= \omega_{2}{}^{2}{}_{0}= \cdots =\omega_{n}{}^{n}{}_{0}\equiv X(t), \quad
\omega_{n+1}{}^{n+1}{}_{0}= \omega_{n+2}{}^{n+2}{}_{0}= \cdots =\omega_{n+m}{}^{n+m}{}_{0}\equiv Y(t), \nonumber \\
&&\mbox{(others)}=0. 
\end{eqnarray}
Under this ansatz non-vanishing components of torsion are
\begin{eqnarray}
T^{i}{}_{,i0}=X-H_1, \quad T^{I}{}_{,I0}=Y-H_2,
\quad \left(H_1\equiv\frac{\dot{a}}{a}, \;\; H_2\equiv\frac{\dot{b}}{b}\right). 
\end{eqnarray} 
By substituting these forms into (\ref{fieldequation}), we obtain following equations:
\begin{eqnarray}
n(\dot{X}+X^2)+m(\dot{Y}+Y^2)=0, \label{feq321} \\
2\dot{X}-\dot{H}_1+(2n-1)X^2+2mXY-(nX+mY)H_1+H_1^2=0, \label{feq322} \\
2\dot{Y}-\dot{H}_2+(2m-1)Y^2+2nXY-(nX+mY)H_2+H_2^2=0, \label{feq323} \\
\ddot{X}+X\dot{H}_1+(nX+mY)\dot{X}+(nX+mY-H_1)XH_1-(n-1)X^3-mX^2Y=0, \label{feq324} \\
\ddot{Y}+Y\dot{H}_2+(nX+mY)\dot{Y}+(nX+mY-H_2)YH_2-(m-1)Y^3-nXY^2=0, \label{feq325}
\end{eqnarray}
where (\ref{feq321}), (\ref{feq322}) and (\ref{feq323}) come from the first equation of (\ref{fieldequation}). On the other hand, (\ref{feq324}) and (\ref{feq325}) come from the second equation of (\ref{fieldequation}). We have five equations here, but it turns out that one of these five is not an independent equation\footnote{By using $\frac{d}{dt}$(\ref{feq321})=0, one can see   
$nX\times(\ref{feq322})+mY\times(\ref{feq323})=n\times(\ref{feq324})+m\times(\ref{feq325})$.}.  
Therefore we regard (\ref{feq321})-(\ref{feq324}) as independent equations to determine $X,Y,H_1$ and $H_2$.
In the paper \cite{Furuta:2006kk}, torsion components ${T^i}_{,i0}$ and ${T^I}_{,I0}$ are identified with a time derivative of some scalar field. Thus if we follow this proposal, additional constraint $X-H_1=Y-H_2$ must be imposed. But here we do not impose this constraint and keep a broader possibility for the solution.

The same as in the previous subsection, the field equations above have the following curvature-less solution:
\begin{eqnarray}
X=Y=0, \quad H_1=\frac{1}{\alpha_1-t}, \;\;H_2=\frac{1}{\alpha_2-t},
\end{eqnarray}
where $\alpha_{1,2}$ are integral constants.

It is difficult to have an analytic result in the case of $X, Y \neq 0$. Here we consider linear approximation of the equations (\ref{feq321})-(\ref{feq324}):
\begin{eqnarray}
\ddot{X}=0, \quad n\dot{X}+m\dot{Y}=0, \nonumber \\
2\dot{X}-\dot{H}_1=0, \quad 2\dot{Y}-\dot{H}_2=0.
 \label{leq}
\end{eqnarray}
These are easily solved by
\begin{eqnarray}
X=\beta t+ \beta', \;\;Y=-\frac{n}{m}\beta t+\beta'', \quad
H_1=2\beta t+\gamma, \;\;H_2=-\frac{2n}{m}\beta t+\tilde{\gamma},
\end{eqnarray}
with integral constants $\beta,\beta',\beta''$ and $\gamma,\tilde{\gamma}$.
Let $\gamma=\tilde{\gamma}=0$, for example. 
Then we have $a(t)\propto \exp( \beta t^2)$ and 
$b(t)\propto \exp(-\frac{n\beta}{m}t^2)$.
If $a(t)$ gives expansion or shrinking in $x^i$ directions, then $b(t)$ gives shrinking or expansion in $y^I$ directions. Acceleration $\ddot{a}$ is positive if $\beta>0$. Acceleration $\ddot{b}$ is negative at early time and becomes positive after $t>m/2n\beta$. Linear approximation may be valid during $t<1/2\beta$. Thus $\ddot{b}$ can be positive if $m/n<1$, within the linear approximation. 

For large $X,Y,H_1$ and $H_2$ we have to take care of full non-linearity of the equations (\ref{feq321})-(\ref{feq324}). Numerical computation is effective to do it. Depending on initial conditions for $(X,Y,H_1,H_2)$, we may experience various periods of evolution; $a(t), b(t)$ give expansion or shrinking with or without acceleration. We left such detailed numerical analysis to future work.

\section{Static Solutions with Spherical Symmetry}
In this section we study solutions which describe static solutions with spherical symmetry. We restrict our interest to (3+1)-dimensional spacetime. The same as in the previous sections, we do not intend to exhaust the most general ansatz which respects spherical symmetry for the spin connection. Instead, we set a simple ansatz in order to investigate the role of torsion. \\

We shall adopt the following ansatz for the metric and the spin connection:
\begin{eqnarray}
&&ds^2=-F(r)^2dt^2+G(r)^2dr^2+r^2(d\theta^2+\sin^2\theta d\phi^2), \nonumber \\
&&\omega_{0}{}^{0}{}_{1} \equiv A(r), \quad 
\omega_{2}{}^{2}{}_{1}=\omega_{3}{}^{3}{}_{1} \equiv B(r), \quad  \omega_{3}{}^{3}{}_{2}=\frac{\cos\theta}{r\sin\theta},
\quad \mbox{(others)}=0.
\end{eqnarray}          
Under this ansatz the non-vanishing components of the torsion are
\begin{eqnarray}
T^{0}{}_{,01}=A-D, \;\;T^{2}{}_{,21}=T^{3}{}_{,31}=B-E, \quad 
\left(D \equiv \frac{F'}{FG}, \;\; E \equiv \frac{1}{rG}\right).
\end{eqnarray}
Substituting these forms into (\ref{fieldequation}), we obtain the following equations:
\begin{eqnarray}
&&\frac{1}{G}(nA'+mB')+nA^2+mB^2=0, \label{feq41}\\
&&\frac{1}{G}(2A'-D')+AD+mAB+mB(A-D)+(A-D)^2=0, \label{feq42} \\
&&\frac{1}{G}(2B'-E')+BE+nAB+nA(B-E)+(B-E)^2+P=0, \label{feq43} \\
&&\left(\frac{1}{G}\partial_r+A+2B-D\right)\left[\frac{1}{G}A'+AD\right]-mAB^2=0, \label{feq44} \\
&&\left(\frac{1}{G}\partial_r+A+2B-E\right)\left[\frac{1}{G}B'+BE\right]-nA^2B+Q=0, \label{feq45} \\
&& \qquad P \equiv B(B-E)+\Big(B^2-\frac{1}{r^2}\Big), \quad Q \equiv B\Big(\frac{1}{r^2}-B^2\Big).
\nonumber 
\end{eqnarray}
We introduced $n=1,m=2$ to write down these equations in more symmetric form. 
Equations (\ref{feq41}), (\ref{feq42}) and (\ref{feq43}) come from the first equation of (\ref{fieldequation}). On the other hand,  (\ref{feq44}) and (\ref{feq45}) come from the second equation of (\ref{fieldequation}).  There are five equations,  however, it turns out that one of them is 
not an independent equation\footnote{One can see $n\times(\ref{feq44})+m\times(\ref{feq45})=0$, using (\ref{feq41}), (\ref{feq42}), (\ref{feq43}) and $\frac{1}{G}\partial_r(\ref{feq41})$.}. Thus we use four equations (\ref{feq41})-(\ref{feq44}) to determine $A,B,D,$ and $E$. In the paper \cite{Furuta:2006kk}, ${T^{0}}_{,01}, {T^{2}}_{,21},$ and ${T^{3}}_{,31}$ are identified with a derivative of some scalar field. If we follow their proposal then additional constraint $A-D=B-E$ must be imposed. Again we do not impose this constraint here in order to keep a broader possibility for the solutions.

\subsection{Analytic approach}
In the torsion-less case, the Schwarzschild spacetime $F^2=1/G^2=1-c_0/r$ is a solution. It corresponds to 
\begin{eqnarray}
A=D=\frac{c_0}{2r^2\sqrt{1-\frac{c_0}{r}}}, \quad 
B=E=\frac{1}{r}\sqrt{1-\frac{c_0}{r}}.
\end{eqnarray}

Similar to the previous sections, we can find a curvature-less solution $A=B=0$. The equations for $D$ and $E$ are
\begin{eqnarray}
\frac{1}{G}D'-D^2=0, \quad 
\frac{1}{G}E'-E^2+\frac{1}{r^2}=0. \label{feq46} 
\end{eqnarray}
Recall that $D=F'/FG$ and $E=1/rG$, then we find two different solutions of (\ref{feq46}). The first solution is 
\begin{eqnarray}
\frac{1}{F}=c'' \left(\int^{r} G(s)ds\right)+c', \quad G= \sqrt{\frac{2}{1+c r^4}}, \quad (A=B=0)
\label{sol42}
\end{eqnarray} 
with integral constants $c,c',$ and $c''$
\footnote{$1/F$ can be written as an elliptic integral. }. 
The second solution is
\begin{eqnarray}
F=\frac{\epsilon'}{\epsilon-r}, \quad G=\frac{1}{\sqrt{2}}, \quad (A=B=0)
\label{sol43}
\end{eqnarray} 
with integral constants $\epsilon$ and $\epsilon'$. These are not an asymptotically flat solution. 

It is difficult to have analytic solution for $A,B\neq 0$ with torsion. Here we consider a linear approximation of the equations (\ref{feq41})-(\ref{feq44}):
\begin{eqnarray}
\frac{1}{G}\partial_r\left( \frac{1}{G}\partial_r A\right)=0, \quad \frac{1}{G}(A'+2B')=0, \nonumber \\
\frac{1}{G}(2A'-D')=0, \quad \frac{1}{G}(2B'-E')=0. 
\end{eqnarray}  
These equations may be valid for the large $r$ region. From these equations, we have
\begin{eqnarray}
\frac{1}{G}A'=-\delta, \quad \frac{1}{G}B'=2\delta,
\quad   \frac{1}{G}D'= -2\delta, \quad \frac{1}{G}E'=4\delta. 
\quad (\delta=const.)
\end{eqnarray} 
 Recalling $E=1/rG$, then the solution is 
 \begin{eqnarray}
&& A=-\delta\int^rG(s)ds, \quad B=2\delta\int^r G(s)ds, \nonumber \\
&&  \log F=-\delta\left(\int^rG(s)ds\right)^2+\delta', \quad 
  \frac{1}{r^2G^2}=(8\delta\log r+\delta'').
 \end{eqnarray}
 Again, this does not give asymptotically flat spacetime. $E$ behaves as $\log r$, so this linear approximation breaks down not only in the small $r$ region
but also in a very large $r$ region.

An asymptotically flat solution can be constructed by a formal power series expansion.
To do this we rewrite (\ref{feq41})-(\ref{feq44}) as first order differential equations:
\begin{eqnarray}
\frac{dA}{dr}&=&\frac{C}{Er}, \nonumber \\
\frac{dB}{dr}&=&-\frac{1}{2} \frac{C+A^2+2B^2}{Er} ,  \nonumber \\
\frac{dC}{dr}&=&\frac{-(3A+2B)C-4A^2B+2AB^2-A^3}{Er}, \nonumber \\
\frac{dD}{dr}&=&\frac{2C+AD+2B(A-D)+(A-D)^2}{Er}, \nonumber \\
\frac{dE}{dr}&=&\frac{-C-A^2+2BE+(B-E)^2-1/r^2}{Er}. \label{firstorder}
\end{eqnarray}
Coefficients of formal $1/r$ expansion can be determined by (\ref{firstorder}).  The result is
\begin{eqnarray}
&&A=\frac{a}{r^2}+\frac{a(3a-2b)}{r^3}+O(r^{-4}), \quad
D=\frac{d}{r^2}+\frac{a^2+3ad-2bd+d^2}{r^3}+O(r^{-4}), \nonumber \\
&&C=\frac{-2a}{r^3}-\frac{a(7a-6b)}{r^4}+O(r^{-5}), \nonumber \\
&&B=\frac{1}{r}-\frac{b}{r^2}+\frac{a^2-8ab+5b^2}{4}\frac{1}{r^3}+O(r^{-4}), \quad  
E=\frac{1}{r}-\frac{a}{r^2}-\frac{13a^2+12ab+b^2}{4}\frac{1}{r^3} +O(r^{-4}). \nonumber \\
\label{powers}
\end{eqnarray}
Constants $a,b,d$ parametrize the solution.  
From this expression we obtain the metric
\begin{eqnarray}
F&=&1-\frac{d}{r}-\frac{d( \frac{a^2}{d}+4a-2b-2d ) }{2r^2}+O(r^{-3}), \nonumber \\
\frac{1}{G}&=&1-\frac{a}{r}-\frac{13a^2-12ab+b^2}{4}\frac{1}{r^2}+O(r^{-3}), 
\end{eqnarray}
and the torsion
\begin{eqnarray}
\hspace{-5mm}
A-D&=&\frac{a-d}{r^2}+\frac{2a^2-2ab-3ad+2bd+d^2}{r^3}+O(r^{-4}), \nonumber \\
B-E&=&\frac{a-b}{r^2}+\frac{7a^2-10ab+3b^2}{2}\frac{1}{r^3}+O(r^{-4}). 
\end{eqnarray}
Although we can formally continue to construct the power series solution to higher order, it is not clear whether the event horizon does appear or not.
From the next subsection we examine a numerical integration to treat full non-linearity of equations and discuss the existence of horizon.   
In the numerical approach, a generic boundary condition does not give an asymptotically flat solution.
Probably the formal power series (\ref{powers}) is not a convergent series and cannot be a full solution.
We may use it as an approximately solution which describes the asymptotic form of a spacetime with some matter in the inner region.

\subsection{Numerical approach}
A numerical approach is effective to treat full non-linearity of coupled differential equations.
For convenience of the analysis, we change variables in differential equations (\ref{firstorder}) as
\begin{eqnarray*}
&&r=\frac{1}{u}, \\
&&A\mapsto u^2A, \;\;D \mapsto u^2D, \quad C \mapsto u^3C, \quad 
B \mapsto uB,\;\; E \mapsto uE.
\end{eqnarray*}  
After this, changing (\ref{firstorder})  becomes
\begin{eqnarray}
\frac{dA}{du}&=&-\frac{1}{u}\left(2A+\frac{C}{E}\right), \nonumber \\
\frac{dB}{du}&=&\frac{1}{u}\left(-B+\frac{B^2}{E}\right)+\frac{1}{2}\frac{C+A^2u}{E}, \nonumber \\
\frac{dC}{du}&=&\frac{1}{u}\left(-3A+\frac{2BC-2AB^2}{E}\right)+\frac{3AC+4A^2B+A^3u}{E}, \nonumber \\
\frac{dD}{du}&=&-\frac{1}{u}\left(2D+\frac{2C+4AB-2BD}{E}\right)+\frac{AD+(A-D)^2}{E}, \nonumber \\
\frac{dE}{du}&=&\frac{1}{u}\left(E+\frac{(B-E)^2-1}{E} \right)+\frac{C-2AB+AE+A^2u}{E}.
\end{eqnarray}
We used the ``desolve" function of Maple to obtain a numerical solution, imposing a  boundary condition at $u=\frac{1}{100}$.  
We have found four  different classes of the numerical solutions which are characterized by singularities\footnote{Here what singularities mean should be understood in a numerical sense, though we expect that it relates to 
a  non-analytic property of solution.}.
Some of them seem to correspond to the event horizon.  We adopt an intuitive criterion for the horizon; if numerically $F(r)/G(r)=0$, then it is identified with the existence of the  horizon,
because $F(r)/G(r)$ is the speed of light seen by an observer at infinity in the case of Schwarzschild spacetime.   
Below we list up solutions we found:

\begin{itemize}
\item Type I : There are two singularities at $u=0$ and $u=u_h$. The solution exists between $0<u<u_h$. 
The point $u=u_h$ is identified with the event horizon.   Around $u=0$, the solution blows up with power laws.

\item Type I{}I : There are two singularities which are identified with horizons. 
The solution exists  between these two horizons.

\item Type I{}I{}I: There is a singularity at $u=u_h$ which is identified with the horizon. 
The solution exists in $u_h<u$.

\item Type I{}V : There is a singularity at $u=0$, but this is not a horizon, 
The solution exists in $0<u$. Around $u=0$, it blows up with power laws.  
\end{itemize}
The curvature tensor $R_{ab}{}^{cd}$ and the scalar curvature $R=R_{ab}{}^{ab}$ can diverge at these horizons. 
However, it could be canceled by divergence of the torsion.
According to the field equation (\ref{fieldequation}), 
a particular scalar combination:
$R+\nabla^c{T^a}_{,ca}+T^{p,qa}T_{p,qa}$ is always zero except at positions of sources.
In this sense the divergences of the curvature tensor and the torsion at these horizons do not immediately mean the singularity of the solution.  

In next subsection, we discuss these solutions in more detail.

\subsection{Details of numerical solutions}
\subsubsection*{type I solution}
An example of type I solution is given by the boundary condition:
\begin{eqnarray}
 (A,B,C,D,E)=\Big(\frac{1}{100},1,\frac{-2}{100},\frac{1}{100},1\Big),  \quad  \mbox{at $u=\frac{1}{100}$} . 
\end{eqnarray}
\begin{figure}
\begin{center}
\rotatebox{-90}{\includegraphics[height=8cm,width=6cm,clip]{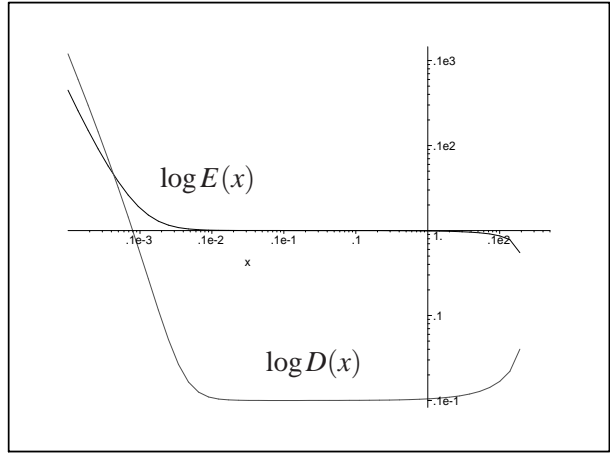}}
\put(-130,-140){$\log D(x)$}
\put(-170,-70){$\log E(x)$}
\end{center}
\caption{Type I solution: Log-Log plot of $D$ and $E$. Horizontal line is $x=\log u$.}
\label{DE1}
\end{figure}
\begin{figure}
\begin{center}
\rotatebox{-90}{\includegraphics[height=6cm, width=5cm,clip]{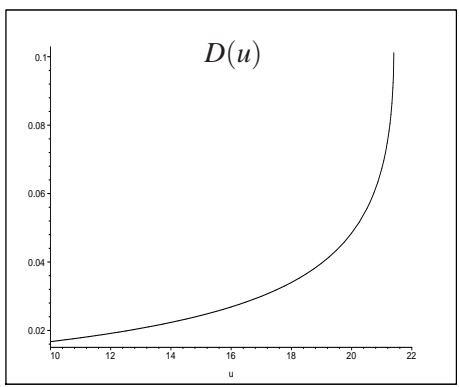}}
\hspace{10mm}
\rotatebox{-90}{\includegraphics[height=6cm,width=5cm,clip]{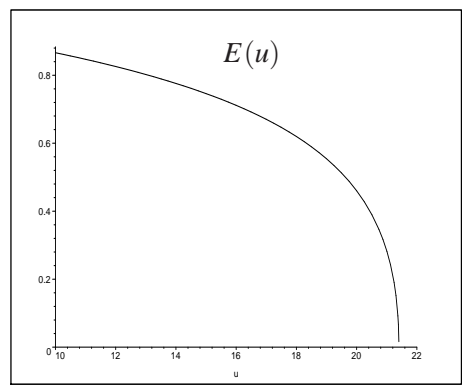}}
\put(-300,-20){$D(u)$}
\put(-90,-20){$E(u)$}
\caption{Type I solution: plots of $D$ and $E$ near the singularity $u_h=21.4$. Horizontal lines are $u$. }
\label{DE1.2}
\end{center}
\end{figure}
Figure \ref{DE1} shows that there are two singularities for solutions. One is located at $u<10^{-20}$. We think that the real position of this singularity is $u=0$, and it is caused from  power law behavior of the solutions around $u\sim 0$. Another singularity is located at $u_h=21.4$. Figure \ref{DE1.2} shows that near $u=u_h$, $D(u)$ and $E(u)$ behave like
\begin{eqnarray}
&&D(u) \sim D_h(u_h-u)^p, \;\;p<0, \nonumber \\ 
&&E(u) \sim E_h(u_h-u)^q, \;\;0<q<1 
\end{eqnarray}
with some positive number $D_h$ and $E_h$
\footnote{Note if $y=(x_0-x)^q$, then $y''<0$ for $0<q<1$.}. Recalling $D(r)=\frac{1}{FG}\frac{dF}{dr}$ and $E(r)=\frac{1}{rG}$, then 
\begin{eqnarray}
F(u) \sim \exp\left(-\frac{{\cal C}}{|p+q|-1}(u_h-u)^{-|p+q|+1} \right), \quad 
\frac{1}{G(u)} \sim (u_h-u)^{q}.
\end{eqnarray}
In the case of  $|p+q|-1\neq 0$, where ${\cal C}$ is some positive numerical factor. If $|p+q|-1>0$, then $F\rightarrow 0$ 
as $u\rightarrow u_h$, while $F$ becomes finite at $u=u_h$ is $|p+q|-1<0$. Then $F/G=0$ at $u_h$.  In the case of $|p+q|=1$, we have $F\sim \log (u_h-u)$. In this case we evaluate $F/G$ as
\begin{eqnarray*}
\left. \frac{F}{G}\right|_{u\rightarrow u_h}
=\left. \left.\frac{1}{G}\right/\frac{1}{F} \right|_{u\rightarrow u_h}  
=\left. \left.\left(\frac{1}{G}\right)'\right/\left(\frac{1}{F}\right)' \right|_{u\rightarrow u_h} 
\propto(u_h-u)^q|_{u\rightarrow u_h}=0.
\end{eqnarray*}
We used $0<q<1$ in the last step.
Now $F/G=0$ at $u=u_h$ has been established. Then we conclude $u_h$ is the event horizon according to our criterion. The type I solution describes the outside of a black hole in an asymptotically non-flat spacetime.

\subsubsection*{type I{}I solution}
An example of the type I{}I solution is given by  boundary condition:
\begin{eqnarray}
 (A,B,C,D,E)=\Big(\frac{1}{100},1,\frac{-2}{100},\frac{1}{100},1-\frac{1}{10}\Big),  \quad  \mbox{at $u=\frac{1}{100}$} . 
\end{eqnarray}
\begin{figure}
\begin{center}
\rotatebox{-90}{\includegraphics[height=6cm, width=5cm,clip]{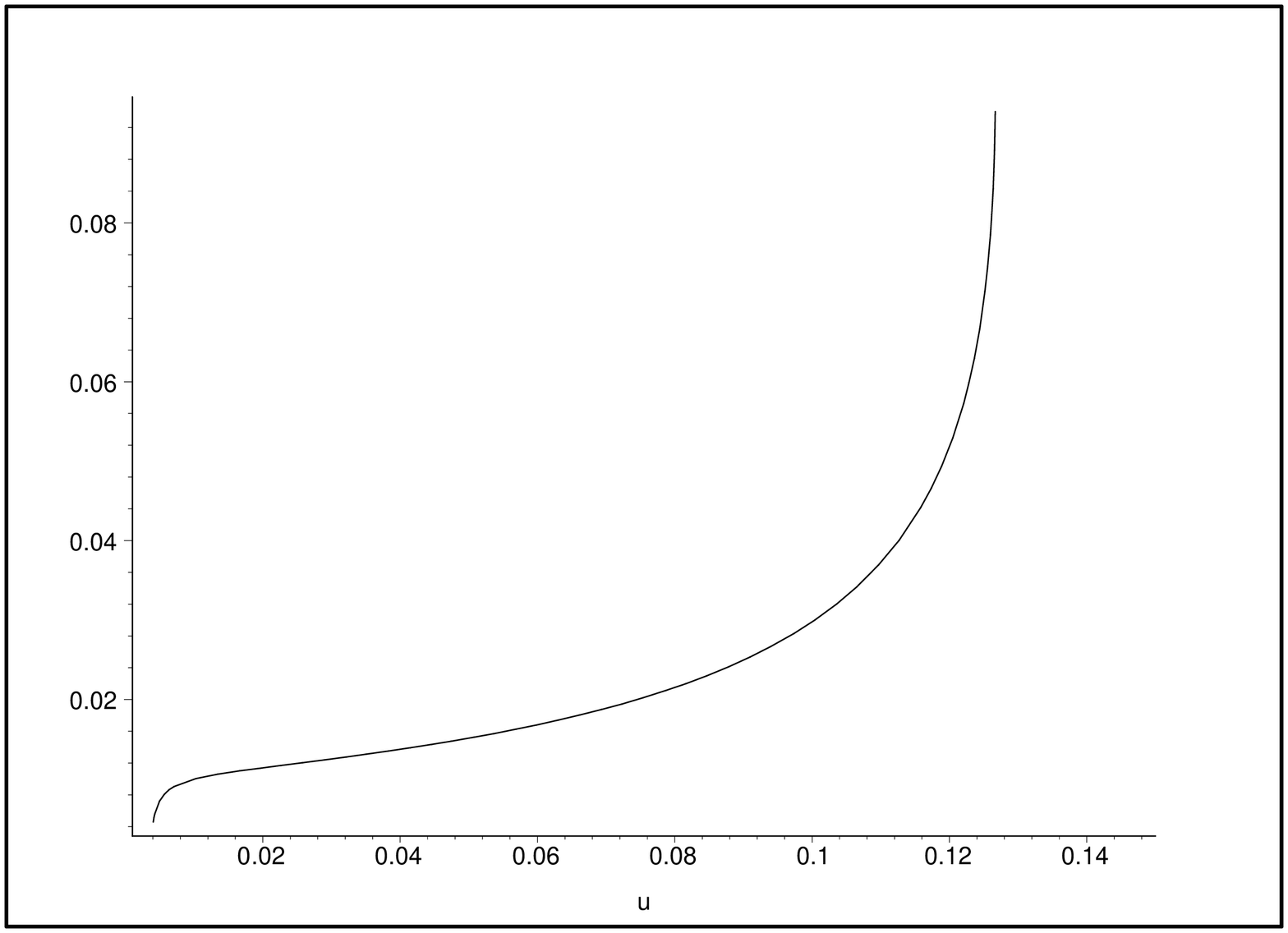}}
\hspace{10mm}
\rotatebox{-90}{\includegraphics[height=6cm,width=5cm,clip]{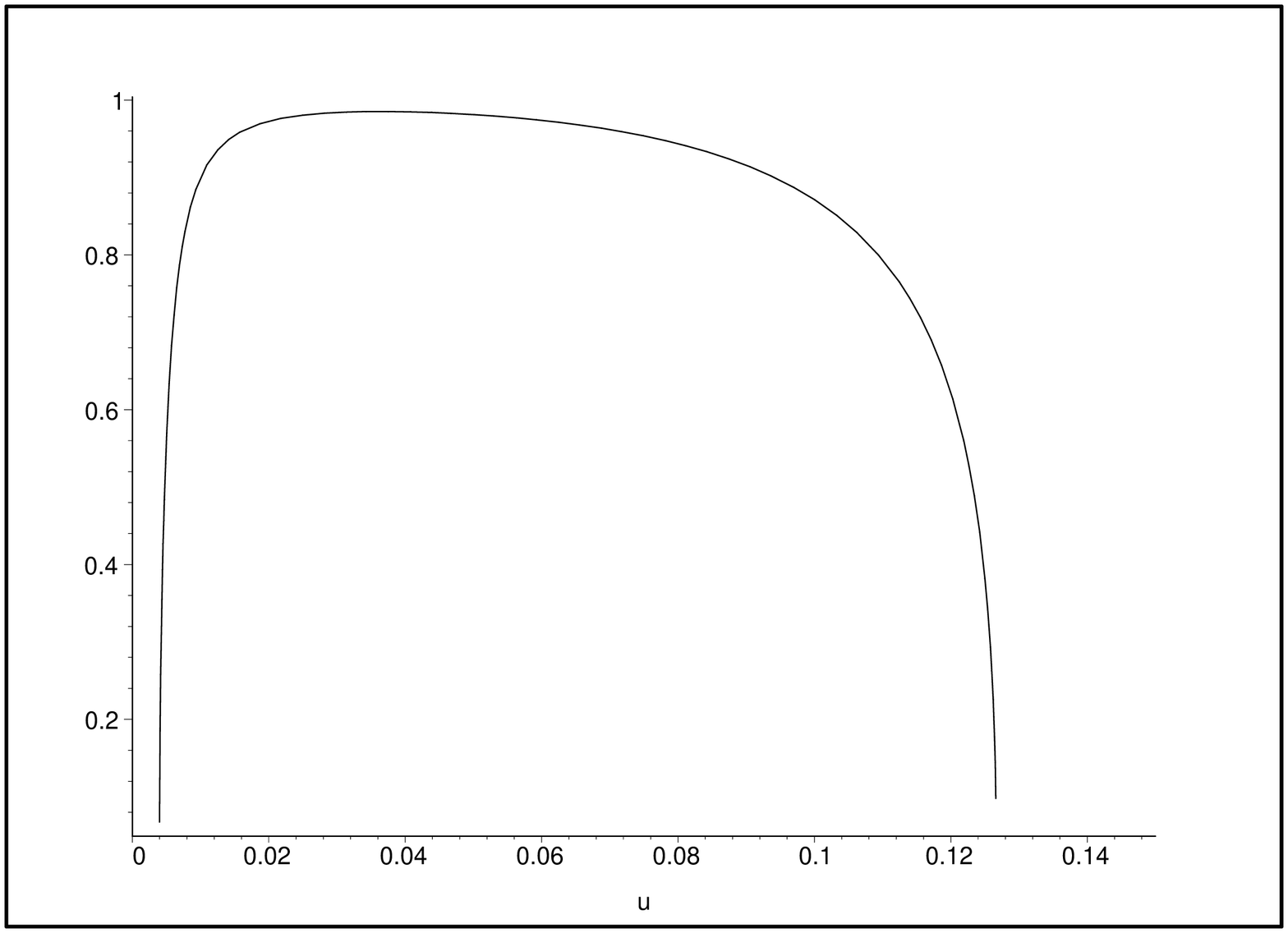}}
\put(-300,-20){$D(u)$}
\put(-70,-20){$E(u)$}
\caption{Type I{}I solution: plots of $D$ and $E$. Horizontal lines are $u$. }
\label{DE2}
\end{center}
\end{figure}
There are two singularities located at $u_{h1}=3.97\times 10^{-3}$ and $u_{h2}=1.27\times 10^{-1}$. Figure \ref{DE2} shows that near $u_{h1}$, $D(u)$ and $E(u)$ behave like 
\begin{eqnarray}
&&D(u) \sim (u-u_{h1})^p, \;\;0<p<1, \nonumber \\ 
&&E(u) \sim (u-u_{h1})^q, \;\;0<q<1, 
\end{eqnarray}
and  near $u_{h2}$, their behavior is  
\begin{eqnarray}
&&D(u) \sim (u_{h2}-u)^p, \;\;p<0, \nonumber \\ 
&&E(u) \sim (u_{h2}-u)^q, \;\;0<q<1 ,
\end{eqnarray}
with some positive numerical factors.
Through an argument similar to that for the type I solution, we conclude that $u_{h1}$ and $u_{h2}$ are the event horizons.

\subsubsection*{type I{}I{}I solution}
An example of the type I{}I{}I solution is given by the boundary condition:
\begin{eqnarray}
 (A,B,C,D,E)=\Big(\frac{1}{100},1-\frac{1}{10},\frac{-2}{100},\frac{1}{100},1-\frac{1}{10}\Big),  \quad  \mbox{at $u=\frac{1}{100}$} . 
\end{eqnarray}
\begin{figure}
\begin{center}
\rotatebox{-90}{\includegraphics[height=6cm, width=5cm,clip]{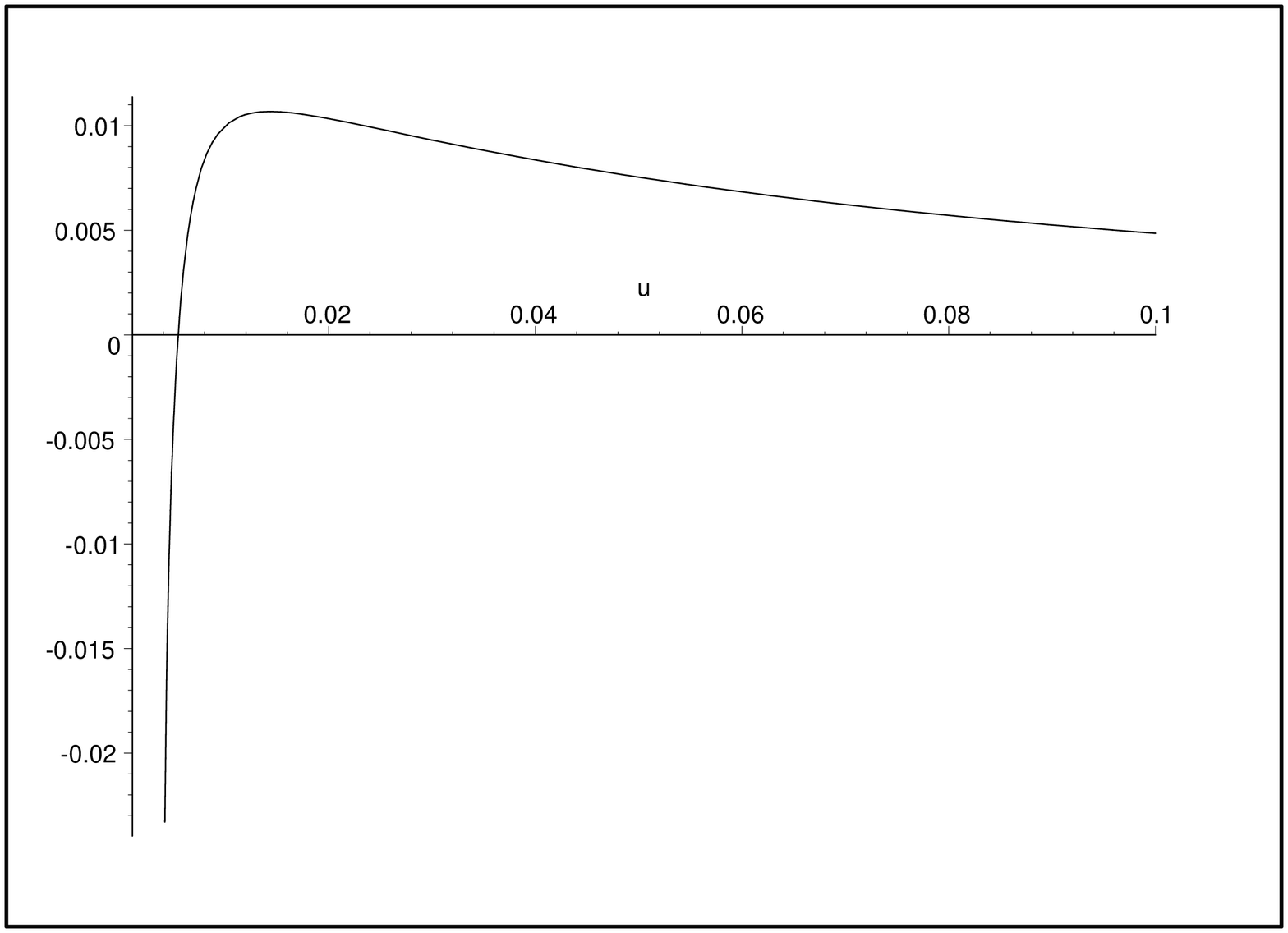}}
\hspace{10mm}
\rotatebox{-90}{\includegraphics[height=6cm,width=5cm,clip]{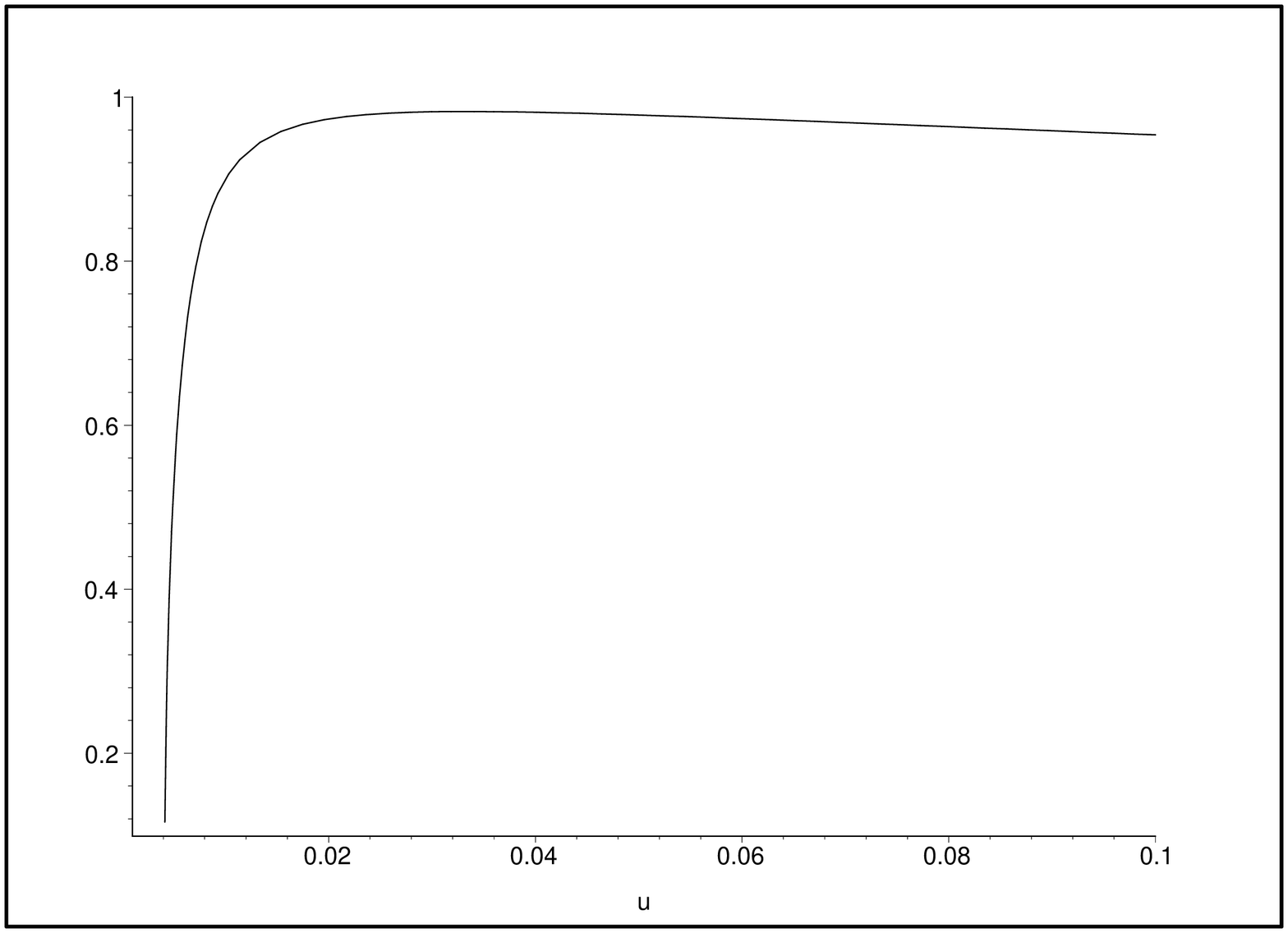}}
\put(-300,-20){$D(u)$}
\put(-70,-30){$E(u)$}
\caption{Type I{}I{}I solution: plots of $D$ and $E$. Horizontal lines are $u$. }
\label{DE3}
\end{center}
\end{figure}
There is a  singularity located at $u_{h}=4.11\times 10^{-3}$. Near $u_{h1}$, $D(u)$ does not have divergent behavior (at least numerically). It implies that $D(u)$ and $E(u)$ in figure \ref{DE3} behave like 
\begin{eqnarray}
&&D(u) \sim -C+D_{h}(u-u_h)^p, \;\;0<p<1, \nonumber \\ 
&&E(u) \sim E_{h}(u-u_h)^q, \;\;0<q<1.
\end{eqnarray}
We conclude that $F/G$ vanishes at $u_h$ using an argument similar to the previous ones. 
Then $u=u_{h}$ is the horizon.

\subsubsection*{type I{}V solution}
An example of the type I{}V solution is given by the  boundary condition:
\begin{eqnarray}
 (A,B,C,D,E)=(\frac{1}{100},1-\frac{1}{10},\frac{-2}{100},\frac{1}{100},1+\frac{1}{10}),  \quad  \mbox{at $u=\frac{1}{100}$} . 
\end{eqnarray}
\begin{figure}
\begin{center}
\rotatebox{-90}{\includegraphics[height=6cm, width=5cm,clip]{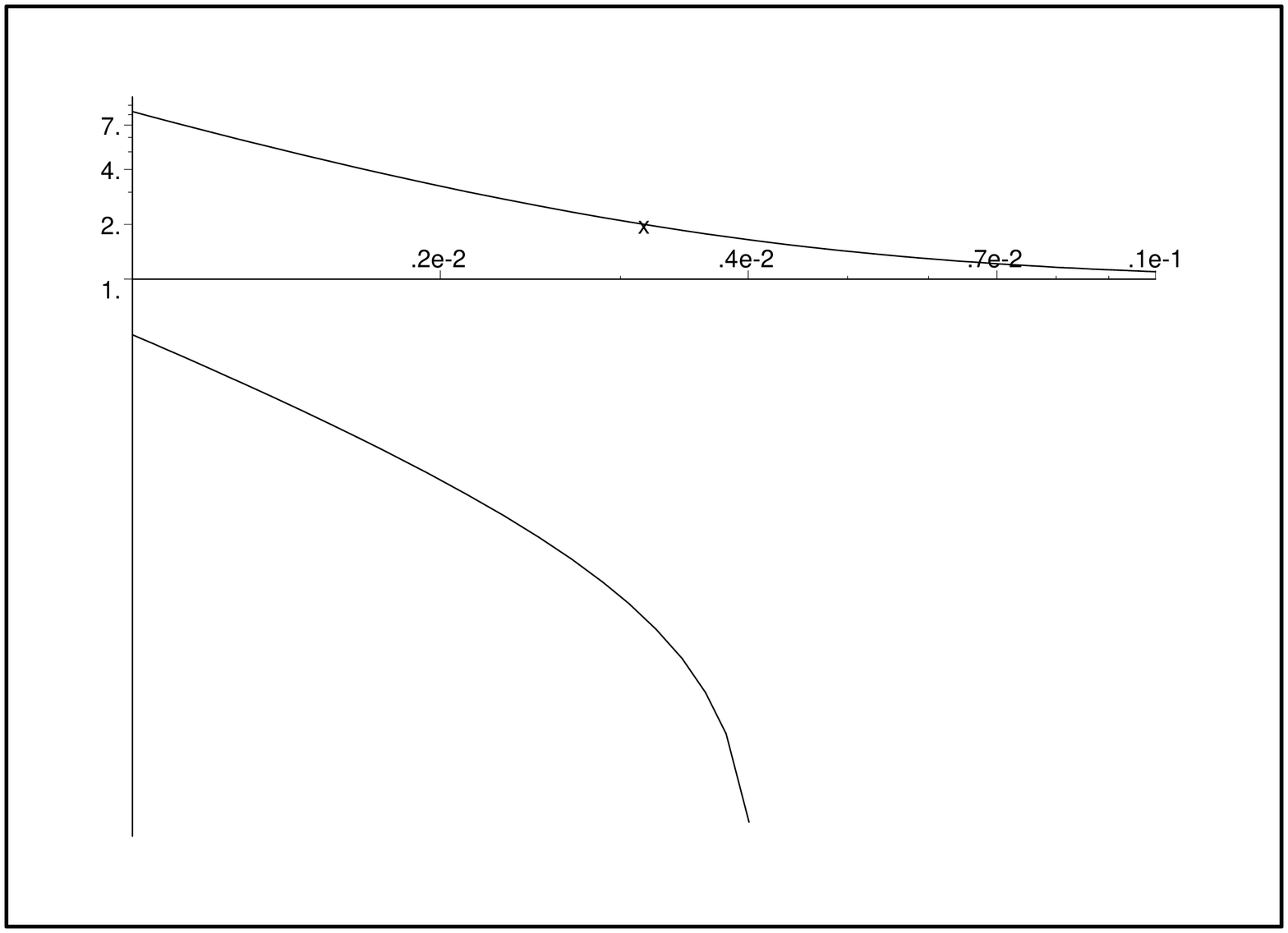}}
\hspace{10mm}
\rotatebox{-90}{\includegraphics[height=6cm,width=5cm,clip]{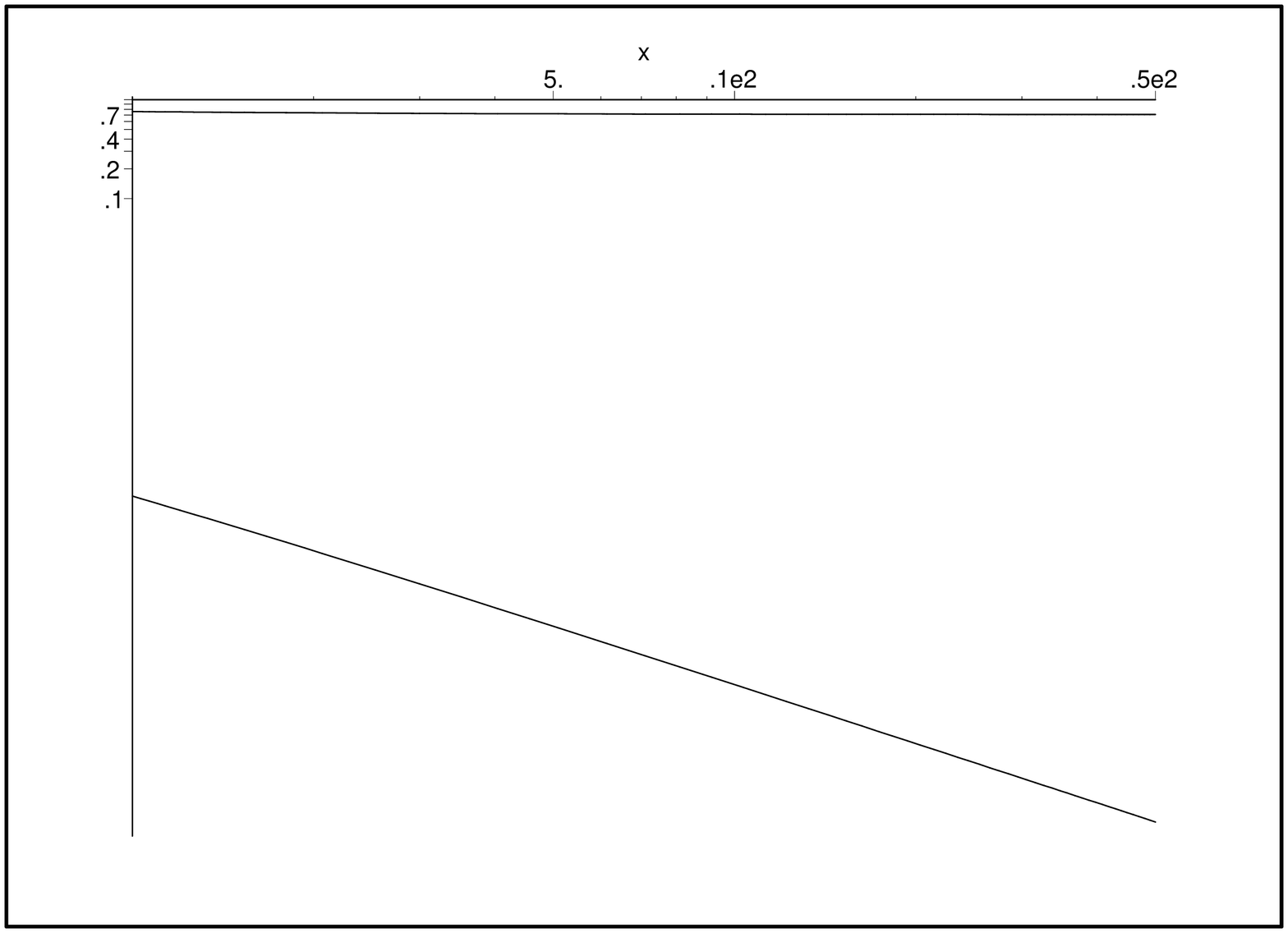}}
\put(-350,-100){$\log (-D(x))$}
\put(-300,-20){$\log E(x)$}
%
\put(-150,-100){$\log D(x)$}
\put(-120,-30){$\log E(x)$}
\caption{Type I{}V solution: plots of $\log D$ and $\log E$, in small and large $u$ regions. 
Horizontal lines are $x=\log u$. 
$D(u)$ becomes to negative for $u<4\times 10^{-3}$.}
\label{DE4}
\end{center}
\end{figure}
Numerical plots of this solution are displayed in figure \ref{DE4}. There is  a singularity located at $u<10^{-20}$. We think that the real singularity is located at $u=0$, 
which is caused by the power law behavior of $D$ and $E$ near $u=0$. There are no other singularities.
Thus we interpret this type I{}V as the solution which describes a spacetime without black hole. \\

Finally we point out that there are similarities between  some of numerical solutions in this subsection and the analytic solution (\ref{sol42}).
In fact the type I{}I{}I solution is  similar to (\ref{sol42}) with $c<0$,  
and  the type I{}V solution is  similar to (\ref{sol42}) with $c>0$.

\section{Summary and Discussion}
We have studied classical solutions of the torsion gravity in empty spacetime, whose field equation stems from the matrix equation of motion of the IKKT model by using a HKK interpretation. The PGT is one of the known torsion gravities which is formulated by a Lagrangian of the vielbein and the spin connection. Our field equation is not derived from such Lagrangian, so it is different from the PGT field equation, even though a matrix equation of motion itself is derived from a matrix model action. Because of the lack of the gravitational Lagrangian, no procedure for direct quantization of the vielbein and the spin connection is  known in our approach. If their quantization rule is the same as the ordinary one, there is an anxiety about negative norm states caused from time-components of these fields as discussed in \cite{Furuta:2006kk}. In this paper we treat these gravitational fields and their field equations as purely classical objects to describe semi-classical aspects of the large $N$ matrix model.      

Time dependent solutions with homogeneity and isotropy were investigated in section 3. We found that introducing torsion provides a simple mechanism to realize an accelerating expansion of spacetime. In PGT torsion degrees of freedom have been used to study cosmological problems like dark energy and inflation (see \cite{Minkevich:2006kq}, for example). To carry out further investigation along this line, we need to take care of contributions corresponding to the energy-momentum tensor from matter, which is beyond our scope in this paper.

Time independent solutions with spherical symmetry were investigated in section 4, with both analytical and numerical methods. We found solutions with singularities, which relate to the event horizons according to our criterion. Except for the formal power series solution, all solutions with the torsion are not asymptotically flat. The asymptotic behavior is power law. It is contrast to PGT, in which an asymptotically flat solution has been found analytically \cite{Bakler:1980fe} (several asymptotically non-flat solutions were also found \cite{Bakler:1983bm}). This would imply that torsions in our study are constrained to be very small in physically acceptable solutions, in order to well describe the solar system in present day\footnote{Actually torsions in the standard model extension are constrained to be order $10^{-31}$ Gev by measurement \cite{Kostelecky:2007kx}.}. However this observation itself does not prohibit a possibility of finding some solutions with a large effect of torsion in past or future period, which describes possible evolutions of the universe .    
More precise investigation on the structure of these solution and their  systematic classification are future problems.

\section*{Acknowledgment}

We would like to thank 
Yosuke Imamura and Chih-Wei Wang
for discussions and comments.
We also thank to Neil Russell for a comment about constraints on torsion.

\appendix
\section{Bianchi identities}
We use differential forms
$T^c=\frac{1}{2}{T_{ab}}^c\theta^a\wedge\theta^b$ and 
${R^c}_d=\frac{1}{2}{{R_{ab}}^c}_d\theta^a\wedge\theta^b$
with  $\theta^a=e^a_{\mu}dx^{\mu}$. 
Then (\ref{defofTR}) can be written as the following form:
\begin{eqnarray}
T^a&=&d\theta^a-{\omega_b}^a\wedge\theta^b, 
\label{torsion2}\\
{R^a}_b&=&d{\omega^a}_b+{\omega^a}_c\wedge{\omega_b}^c
\label{curvature2}
\end{eqnarray}
where exterior derivative $d$ acts as $dF=dx^\mu {e_{\mu}}^a\partial_{a}\wedge F$ on a differential form $F$. We have introduced a one form ${\omega_a}^b={\omega_{c,a}}^b\theta^c$. 
To obtain (\ref{torsion2}), we also used a property $e^{\mu}_b\partial_{a}e_{\mu}^c=-e_{\mu}^c\partial_ae_{b}^{\mu}$
which follows from ${e^c}_{\mu}{e^{\mu}}_b={\delta^c}_b$. 

Now we may drive Bianchi identities using (\ref{torsion2}) and (\ref{curvature2}).
First we take the exterior derivative of (\ref{torsion2}), then obtain 
\begin{eqnarray}
0&=&d T^d +d{\omega_X}^d\wedge\theta^X-{\omega_X}^d\wedge d\theta^X \nonumber \\
&=&\frac{1}{2}(
\partial_X{T_{YZ}}^d\;\theta^X\wedge\theta^Y\wedge\theta^Z
+{T^d}_{YZ}\;d\theta^Y\wedge\theta^Z
-{T^d}_{YZ}\;\theta^Y\wedge d\theta^Z)
+d{\omega_X}^d\wedge\theta^X-{\omega_X}^d\wedge d\theta^X \nonumber \\
\label{dtorsion2}
\end{eqnarray}
Using (\ref{torsion2}) and (\ref{curvature2}), then (\ref{dtorsion2}) becomes
\begin{eqnarray*}
0&=&\frac{1}{2}\partial_X{T^d}_{YZ}\;\theta^X\wedge\theta^Y\wedge\theta^Z 
+\frac{1}{2}{T^d}_{YZ}({\omega_S}^Y\theta^S\wedge \theta^Z-\theta^Y\wedge {\omega_S}^Z\theta^S)-{\omega_S}^d\wedge T^S \nonumber \\
&&+\frac{1}{2}{T^d}_{YZ}(T^Y\wedge\theta^Z-\theta^Y\wedge T^Z) \nonumber \\
&&+{R_S}^d\wedge\theta^S \nonumber \\
&&-{\omega_X}^S\wedge{\omega_S}^d\wedge\theta^X-{\omega_X}^d\wedge{\omega_S}^X\wedge\theta^S \nonumber \\
&=&\frac{1}{2}(
\partial_a{T^d}_{bc}+{T^d}_{Yc}{\omega_{a,b}}^Y-{T^d}_{aZ}{\omega_{b,c}}^Z-{\omega_{a,S}}^d{T^S}_{bc}
+{T^d}_{Yc}{T^Y}_{ab}+{R_{abc}}^d
)\theta^a\wedge\theta^b\wedge\theta^c  \nonumber \\
&=&\frac{1}{2}(\nabla_a{T^d}_{bc}+{T^d}_{Yc}{T^Y}_{ab}+{R_{abc}}^d
)\theta^a\wedge\theta^b\wedge\theta^c.
\end{eqnarray*}
In this way we obtained the first Bianchi identity
\begin{eqnarray}
\nabla_{(a}{T^d}_{,bc)}+{T^d}_{,e(c}{T^e}_{,ab)}+{R_{(abc)}}^d=0 \label{bianchi1}.
\end{eqnarray} 
Similarly, from (\ref{curvature2}), we can derive the second Bianchi identity  
\begin{eqnarray}
\nabla_{(a}{{R_{bc)}}^d}_e-{T^f}_{(,ab}{{R_{c)f}}^d}_e=0.
\label{bianchi2}
\end{eqnarray}
Here we see that equations (\ref{jacobi}) which follow from the matrix Jacobi identity are identical with Bianchi identities (\ref{bianchi1}) and (\ref{bianchi2}).
Thus all ${T^c}_{,ab}$ and ${R_{ab}}^{cd}$ defined by (\ref{defofTR}) satisfy (\ref{jacobi}).



\begin{thebibliography}{99}
\bibitem{Ishibashi:1996xs}
  N.~Ishibashi, H.~Kawai, Y.~Kitazawa and A.~Tsuchiya,
  ``A large-N reduced model as superstring,''
  Nucl.\ Phys.\  B {\bf 498} (1997) 467
  [arXiv:hep-th/9612115].

\bibitem{Banks:1996vh}
  T.~Banks, W.~Fischler, S.~H.~Shenker and L.~Susskind,
  ``M theory as a matrix model: A conjecture,''
  Phys.\ Rev.\  D {\bf 55} (1997) 5112
  [arXiv:hep-th/9610043].

\bibitem{Hanada:2005vr}
  M.~Hanada, H.~Kawai and Y.~Kimura,
  ``Describing curved spaces by matrices,''
  Prog.\ Theor.\ Phys.\  {\bf 114} (2006) 1295
  [arXiv:hep-th/0508211].
  
\bibitem{Kawai:2007zz}
  H.~Kawai,
  ``Curved space-times in matrix models,''
  Prog.\ Theor.\ Phys.\ Suppl.\  {\bf 171} (2007) 99.

\bibitem{Hanada:2006gg}
  M.~Hanada, H.~Kawai and Y.~Kimura,
  ``Curved superspaces and local supersymmetry in supermatrix model,''
  Prog.\ Theor.\ Phys.\  {\bf 115} (2006) 1003
  [arXiv:hep-th/0602210].

\bibitem{Saitou:2006ca}
  T.~Saitou,
  ``Bosonic massless higher spin fields from matrix model,''
  JHEP {\bf 0606} (2006) 010
  [arXiv:hep-th/0604103].
  
  T.~Saitou,
  ``Superfield formulation of 4D, N=1 massless higher spin gauge field   theory
  and supermatrix model,''
  JHEP {\bf 0707} (2007) 057
  [arXiv:0704.2449 [hep-th]].
 

\bibitem{Hanada:2006ei}
  M.~Hanada,
  ``Regularization of the covariant derivative on curved space by finite
  matrices,''
  Prog.\ Theor.\ Phys.\  {\bf 115} (2006) 1189
  [arXiv:hep-th/0606163].

\bibitem{Furuta:2006kk}
  K.~Furuta, M.~Hanada, H.~Kawai and Y.~Kimura,
  ``Field equations of massless fields in the new interpretation of the matrix
  model,''
  Nucl.\ Phys.\  B {\bf 767} (2007) 82
  [arXiv:hep-th/0611093].

\bibitem{Matsuo:2008yd}
  T.~Matsuo, D.~Tomino, W.~Y.~Wen and S.~Zeze,
  ``Quantum gravity equation in large N Yang-Mills quantum mechanics,''
  JHEP {\bf 0811} (2008) 088
  [arXiv:0807.1186 [hep-th]].


\bibitem{Steinacker:2007dq}
  H.~Steinacker,
  ``Emergent Gravity from Noncommutative Gauge Theory,''
  JHEP {\bf 0712} (2007) 049
  [arXiv:0708.2426 [hep-th]].

  H.~S.~Yang and M.~Sivakumar,
  ``Emergent Gravity from Quantized Spacetime,''
  arXiv:0908.2809 [hep-th].



\bibitem{Hehl:1976kj}
  F.~W.~Hehl, P.~Von Der Heyde, G.~D.~Kerlick and J.~M.~Nester,
  ``General Relativity With Spin And Torsion: Foundations And Prospects,''
  Rev.\ Mod.\ Phys.\  {\bf 48} (1976) 393.

\bibitem{Hammond:2002rm}
  R.~T.~Hammond,
  ``Torsion Gravity,''
  Rept.\ Prog.\ Phys.\  {\bf 65} (2002) 599.



\bibitem{Gonner:1984rw}
  H.~Gonner and F.~Mueller-Hoissen,
  ``Spatially Homogeneous And Isotropic Spaces In Theories Of Gravitation With
  Torsion,''
  Class.\ Quant.\ Grav.\  {\bf 1} (1984) 651.



\bibitem{Minkevich:2006kq}
  A.~V.~Minkevich, A.~S.~Garkun and V.~I.~Kudin,
  ``Homogeneous isotropic cosmological models with pseudoscalar torsion
  function in Poincare gauge theory of gravity and accelerating universe,''
  arXiv:gr-qc/0612116.

  K.~F.~Shie, J.~M.~Nester and H.~J.~Yo,
  ``Torsion Cosmology and the Accelerating Universe,''
  Phys.\ Rev.\  D {\bf 78} (2008) 023522
  [arXiv:0805.3834 [gr-qc]].


\bibitem{Bakler:1980fe}
  P.~Bakler,
  ``The Unique Spherically Symmetric Solution Of The U(4) Theory Of Gravity In
  Phys.\ Lett.\  B {\bf 94} (1980) 44.

\bibitem{Bakler:1983bm}
  P.~Bakler,
  ``Spherically Symmetric Solutions Of The Poincare Gauge Field Theory,''
  Phys.\ Lett.\  A {\bf 96} (1983) 279.


\bibitem{Kostelecky:2007kx}
  V.~A.~Kostelecky, N.~Russell and J.~Tasson,
  ``New Constraints on Torsion from Lorentz Violation,''
  Phys.\ Rev.\ Lett.\  {\bf 100} (2008) 111102
  [arXiv:0712.4393 [gr-qc]].




\end{thebibliography}
\end{document}